\documentclass[aps,twocolumn,superscriptaddress,nofootinbib,tightenlines,
preprintnumbers,showkeys,floatfix]{revtex4-1}

\usepackage[utf8]{inputenc}
\usepackage[english]{babel}
\usepackage{amssymb,amsthm,amsmath,amstext,amsbsy,amsopn}
\usepackage{bbm}
\usepackage{nicefrac}
\usepackage{slashed}
\usepackage{graphicx}
\usepackage{hyperref}
\usepackage{leftidx}
\usepackage{environ}
\usepackage{mathtools}
\usepackage{xspace}
\usepackage{array}
\usepackage{xcolor}
\usepackage{pgfplotstable}
\usepackage{isotope}

\newcommand{\ie}{{i.e.}\xspace}
\newcommand{\eg}{{e.g.}\xspace}

\newcommand{\abinitio}{\textit{ab initio}\xspace}

\newcommand{\MeV}{\ensuremath{\mathrm{MeV}}}

\newcommand{\OO}{\mathcal{O}}

\newcommand{\ket}[1]{|#1\rangle}

\newcommand*\rvec[1]%
{\ensuremath{\overset{\smash{\raisebox{-1.5pt}{\tiny$\rightarrow$}}}{#1}}}
\newcommand*\lvec[1]%
{\ensuremath{\overset{\smash{\raisebox{-1.5pt}{\tiny$\leftarrow$}}}{#1}}}

\NewEnviron{subalign}[1][]{%
\begin{subequations}\begin{align}
  \BODY
\end{align}\label{#1}\end{subequations}
}

\NewEnviron{spliteq}{%
\begin{equation}\begin{split}
  \BODY
\end{split}\end{equation}
}

\newcolumntype{K}[1]{>{\centering\arraybackslash}p{#1}}

\newcommand{\Nmax}{\ensuremath{N_\text{max}}}
\newcommand{\hw}{\ensuremath{\hbar\Omega}}

\newcommand{\NEC}{\ensuremath{N_\text{EC}}\xspace}
\newcommand{\NLEC}{\ensuremath{N_\text{LEC}}\xspace}
\newcommand{\NMV}{\ensuremath{N_\text{mv}}\xspace}

\renewcommand{\vec}[1]{\mathbf{#1}}

\begin{document}

\title{Eigenvector Continuation as an Efficient and Accurate Emulator for\\
Uncertainty Quantification}

\author{S.~K\"onig}
\email{skoenig@ncsu.edu}
\affiliation{Institut für Kernphysik, Technische Universität Darmstadt,
64289 Darmstadt, Germany}
\affiliation{ExtreMe Matter Institute EMMI,
GSI Helmholtzzentrum für Schwerionenforschung GmbH,
64291 Darmstadt, Germany}
\affiliation{Department of Physics, North Carolina State University,
Raleigh, NC 27695, USA}

\author{A.~Ekstr\"om}
\email{andreas.ekstrom@chalmers.se}
\affiliation{Department of Physics, Chalmers University of Technology, SE-412
96 G\"oteborg, Sweden}

\author{K.~Hebeler}
\email{hebeler@theorie.ikp.physik.tu-darmstadt.de}
\affiliation{Institut für Kernphysik, Technische Universität Darmstadt,
64289 Darmstadt, Germany}
\affiliation{ExtreMe Matter Institute EMMI,
GSI Helmholtzzentrum für Schwerionenforschung GmbH,
64291 Darmstadt, Germany}

\author{D.~Lee}
\email{leed@frib.msu.edu}
\affiliation{Facility for Rare Isotope Beams \& Department of Physics
and Astronomy, Michigan State University, MI 48824, USA}

\author{A.~Schwenk}
\email{schwenk@physik.tu-darmstadt.de}
\affiliation{Institut für Kernphysik, Technische Universität Darmstadt,
64289 Darmstadt, Germany}
\affiliation{ExtreMe Matter Institute EMMI,
GSI Helmholtzzentrum für Schwerionenforschung GmbH,
64291 Darmstadt, Germany}
\affiliation{Max-Planck-Institut f\"ur Kernphysik,
Saupfercheckweg 1,
69117 Heidelberg, Germany}

\begin{abstract}
First principles calculations of atomic nuclei based on microscopic nuclear
forces derived from chiral effective field theory (EFT) have blossomed in the
past years.
A key element of such \abinitio studies is the understanding and quantification
of systematic and statistical errors arising from the omission of higher-order
terms in the chiral expansion as well as the model calibration.
While there has been significant progress in analyzing theoretical uncertainties
for nucleon-nucleon scattering observables, the generalization to multi-nucleon
systems has not been feasible yet due to the high computational cost of
evaluating observables for a large set of low-energy couplings.
In this Letter we show that a new method called eigenvector continuation (EC)
can be used for constructing an efficient and accurate emulator for nuclear
many-body observables, thereby enabling uncertainty quantification in
multi-nucleon systems.
We demonstrate the power of EC emulation with a proof-of-principle calculation
that lays out all correlations between bulk ground-state observables in the
few-nucleon sector.
On the basis of \abinitio calculations for the ground-state energy and radius in
\isotope[4]{He}, we demonstrate that EC is more accurate and efficient compared
to established methods like Gaussian processes.
\end{abstract}

\maketitle

\paragraph*{Introduction}

In recent years significant progress has been achieved in the theoretical and
algorithmic development of sophisticated many-body methods that allow the study
of atomic nuclei up to mass number $A \simeq 100$ (see, \eg,
Refs.~\cite{Hagen:2013nca,Hebe15ARNPS,Hergert:2015awm,Tichai:2018mll,
Barb17SCGFlectnote,Morr17Tin} and references therein) based on nucleon-nucleon
(NN) and three-nucleon (3N) interactions derived from chiral
EFT~\cite{Epelbaum:2008ga,Machleidt:2011zz,Hamm13RMP,Hebeler:2020ocj}.
Given these many-body advances, the development of novel and more accurate
nuclear interactions is a very active field of research.
In addition to the theoretical work towards understanding how nuclei emerge from
EFTs of the strong interaction, much effort is spent on the calibration of model
parameters, \eg, low-energy constants (LECs) in EFT descriptions of
nuclear interactions. In principle,
calculations based on such interactions allow for a rigorous
quantification of theoretical uncertainties stemming both from the
parameter-estimation procedure as well as from truncating the EFT
expansion at a given order.
A rigorous uncertainty analysis is certainly possible and requires a careful
determination of relevant
covariances~\cite{Ekstrom:2014dxa,NavarroPerez2014,Carlsson:2015vda} and
subsequent error propagation in all model predictions.
Recently, Bayesian inference has been identified as a powerful and versatile
tool for statistical analysis of EFTs, see for example
Refs.~\cite{Furnstahl:2014xsa,Furnstahl:2015rha,Wesolowski:2015fqa,%
Perez:2015ufa,Zhang:2015ajn,Melendez:2017phj,Wesolowski:2018lzj,%
Ekstrom:2019twv}.

Both parameter estimation and the calculation of posterior probability
distributions for nuclear EFT or model predictions typically require extensive
numerical sampling in a high-dimensional parameter space.
Except for the simple two-nucleon sector, repeated calculation of nuclear
many-body observables quickly becomes prohibitively expensive to allow for
sample sizes sufficiently large to be meaningful.
This work presents a solution to overcome this obstacle.

There are clear indications that many-body observables
contain useful information for calibrating nuclear forces.
For example, a fit of LECs to nuclear data including binding energies and radii
of selected oxygen and carbon isotopes~\cite{Ekstrom:2015rta} showed that
exploiting the information content of complex observables is phenomenologically
important.
In a similar spirit, input from $\alpha$-$\alpha$ scattering data has been used
to constrain two-nucleon forces~\cite{Elhatisari:2015iga,Elhatisari:2016owd}.
In addition, it is clear that at least three-nucleon forces are
necessary for an accurate theoretical description of nuclear systems based on
EFT interactions.
The LECs that enter for multi-nucleon forces need to be determined using
calculations of light nuclei (typically $A=3,4$ are used), and already such
calculations can incur a significant computational cost when a large number of
them is needed.

\begin{figure*}[t!]
\centering
\begin{minipage}{0.5\textwidth}
\centering
\includegraphics[width=0.99\columnwidth]{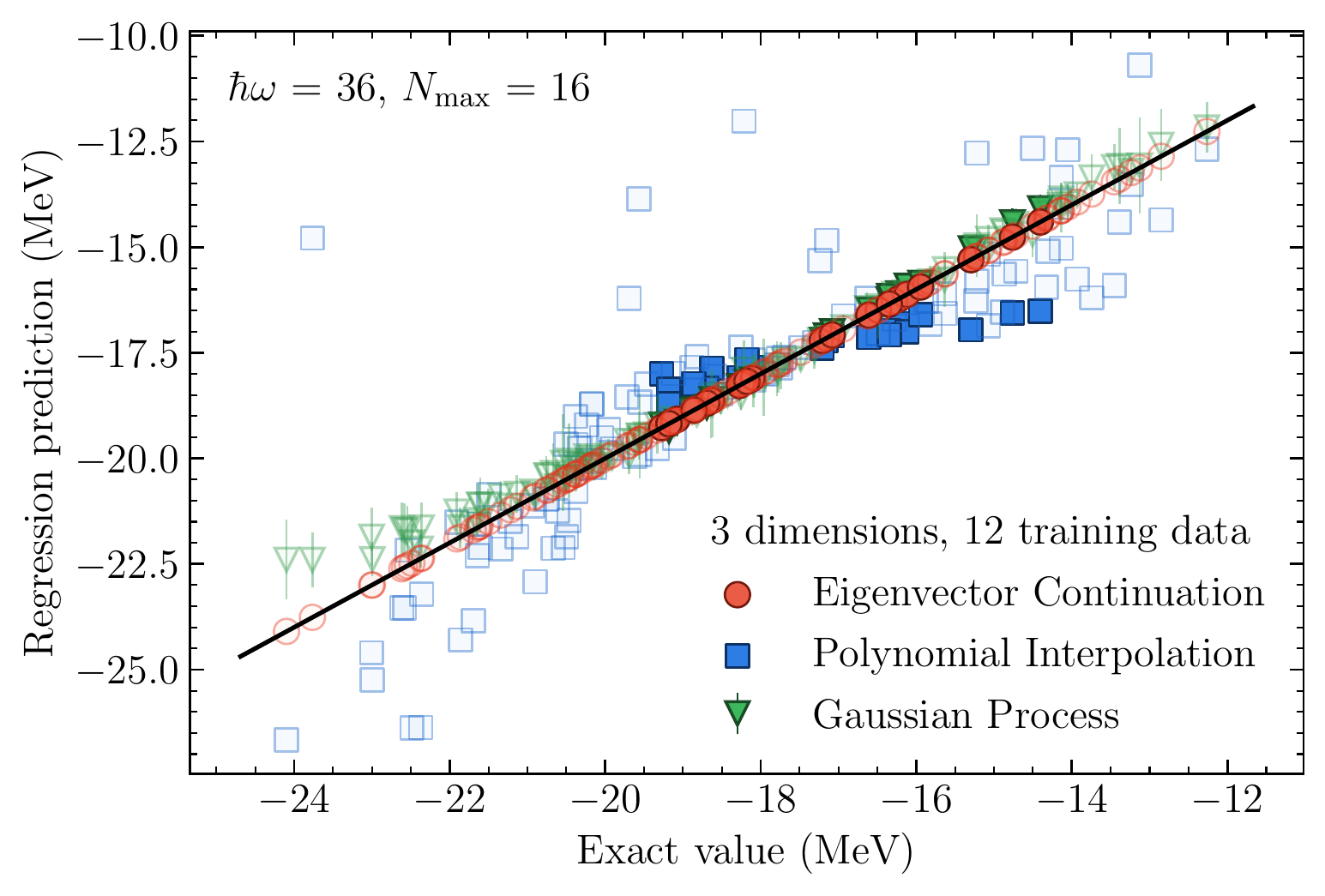}
\end{minipage}%
\begin{minipage}{0.5\textwidth}
\centering
\includegraphics[width=0.99\columnwidth]{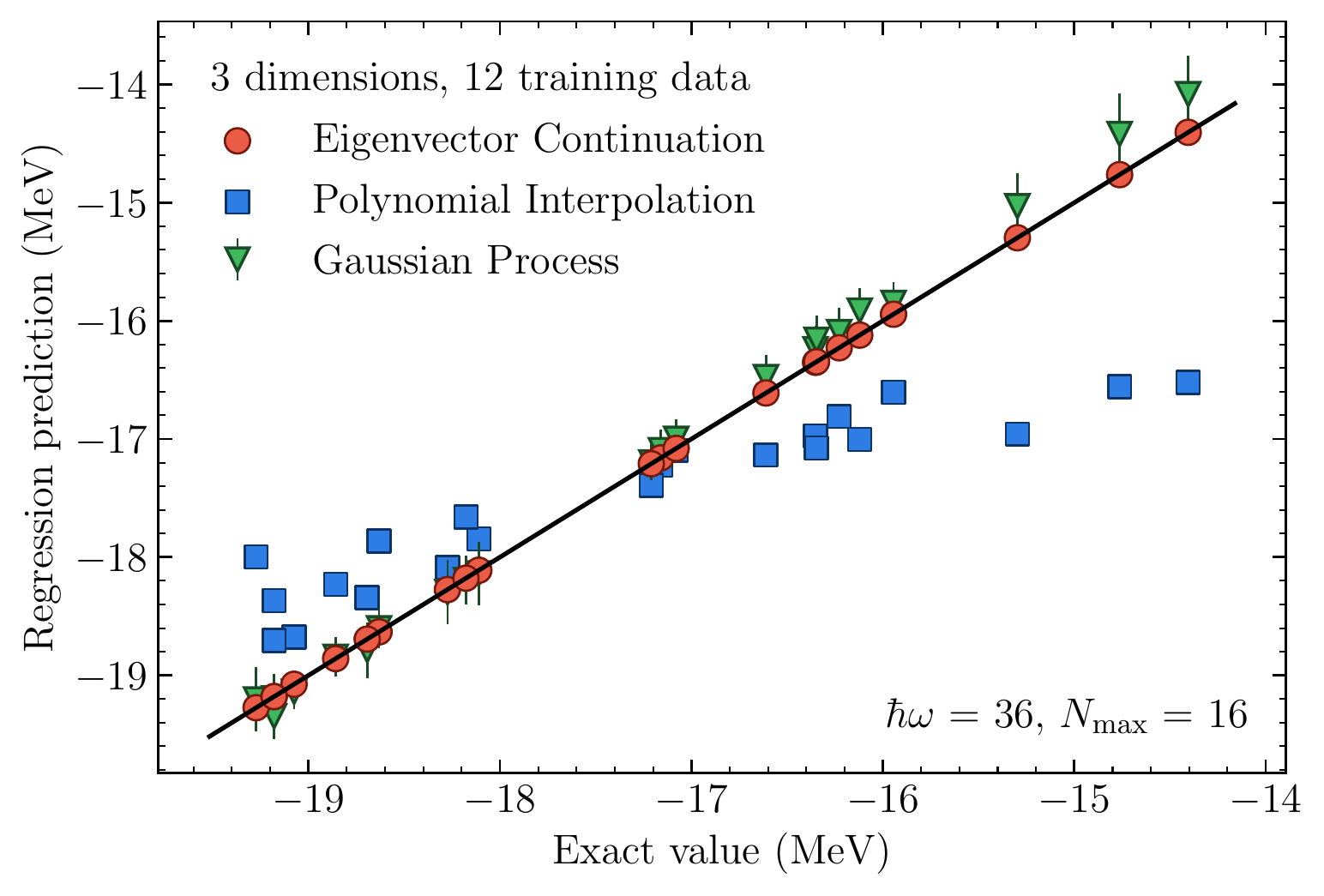}
\end{minipage}
\caption{%
Comparison of different emulators for the \isotope[4]{He} ground-state energy
using 12 training data points to explore a space where three LECs are varied.
The left panel includes samples for both interpolation (solid symbols) and
extrapolation (semi-transparent symbols).
See main text on how these are
defined.
The right panel shows the same data restricted to interpolation
samples (note the smaller axis range).}
\label{fig:Xval_4He_E_Nmax4_hw28-pge-d3-n12}
\end{figure*}

This significant computational cost highlights the importance of developing fast
and accurate methods that make it possible to sample large parameter spaces
using emulators, \ie, calculations that sacrifice the accuracy of an exact
calculation for a significant gain in speed.
The simplest such method, polynomial interpolation between a set of points
within the parameter space, is usually not a viable option for a lack of both
accuracy and efficiency.
Gaussian processes (GP)~\cite{Rasmussen:2006} are useful for leveraging
expensive statistical analyses in nuclear
theory~\cite{Ekstrom:2019twv,Neufcourt}.
As a machine-learning method they can be advantageous for systematically
exploring large parameter spaces and by design provide uncertainties of the
emulator output, but like polynomials they are still limited to interpolation
within a set of training data and cannot be used for reliable extrapolations.
In this Letter, we explore eigenvector continuation (EC), introduced in
Ref.~\cite{Frame:2017fah}, as an alternative to overcome this limitation while
at the same time being significantly more accurate than GP at reduced
numerical cost.
We find that EC performs accurate extrapolations in multi-dimensional parameter
domains even to points far outside the training data set used to construct the
emulator, and that it provides a significantly more efficient and accurate
emulator of nuclear systems than a Gaussian process.

\paragraph*{Formalism}

Eigenvector continuation is based on the fact that when a Hamiltonian depends
smoothly on some real-valued control parameter, any eigenvector of the
Hamiltonian is a smooth function of that parameter as well.
Furthermore, the eigenvector trajectory traced out as the parameter is varied
can be well approximated by a finite-dimensional manifold~\cite{Frame:2017fah}.
This last statement can be turned into a variational method for computing the
eigenvector for any value of the control parameter.

Consider a Hamiltonian $H(c)$ that varies smoothly with real parameter $c$.
The ground-state eigenvector $\ket{v_{0}(c)}$ can be well approximated as some
linear combination of the ground-state eigenvectors
$\ket{v_{0}(c^{[1]})},\cdots, \ket{v_{0}(c^{[N]})}$ at ``training points"
$c^{[1]}, \cdots, c^{[N]}$.
In order to determine the desired linear combination that best approximates
$\ket{v_{0}(c)}$, we simply find the ground state of $H(c)$ projected onto the
subspace spanned by $\ket{v_{0}(c^{[1]})},\cdots, \ket{v_{0}(c^{[N]})}$.
In Ref.~\cite{Frame:2017fah} the applications of EC focused mainly on
extrapolation in cases where the direct calculation of $\ket{v_{0}(c)}$ was not
possible due to computational issues such as the Monte Carlo sign problem.
In this work we will use EC for both interpolation and extrapolation.
We also consider, for the first time, the extension of EC to Hamiltonians that
depend on more than one control parameter.

Specifically, we explicitly demonstrate the advantages of using EC for
constructing a fast and accurate emulator for nonrelativistic calculations of
the \isotope[4]{He} nucleus.
While this application is a benchmark case that is particularly relevant for
nuclear physics, the very general mathematical underpinnings of EC enable the
emulation of expensive problems across several disciplines also outside of
physics provided only that they can be formulated as an eigenvalue problem.
Eigenvector continuation moreover supports full reconstruction of the emulated
eigenvector (wavefunction).
To demonstrate this we consider both the ground-state energy $E$ and radius $r$
of \isotope[4]{He} nucleus, as functions of the 16 LECs $\vec{c}$ in a
particular chiral potential $V(\vec{c})$ for the strong
interaction~\cite{Carlsson:2015vda}, entering the Schr\"odinger equation
$H(\vec{c})\ket{\psi(\vec{c})} = E\ket{\psi(\vec{c})}$.

Training the EC emulator consists of building a basis to span an eigenvector
subspace.
For this we must obtain exact eigenvectors (wavefunctions)
$|\psi(\vec{c}^{[i]})$ for a set of $\NEC$ points $\vec{c}_1, \dots
,\vec{c}_{\NEC}$ across the chosen 16-dimensional parameter domain of the LECs.
We formulate the Schr\"odinger equation for \isotope[4]{He} as an eigenvalue
equation using the no-core shell model (NCSM)~\cite{Barrett:2013nh}.
This is a variational basis-expansion method, also known as ``configuration
interaction'' in quantum chemistry.
The exact wave function of the Hamiltonian $H(\vec{c}_i)$ is expanded in
eigenfunctions of the harmonic-oscillator (HO) potential, yielding a Hamiltonian
represented as a matrix in this HO basis that is subsequently diagonalized.
Considering low-energy states motivates a truncation of this expansion based on
a maximum number of oscillator quanta $\Nmax$.
Another parameter characterizing the basis is the oscillator frequency $\hbar
\Omega$.
For $\Nmax\to\infty$, the choice of frequency is arbitrary, but for each
truncated basis there is a residual dependence of results on $\hbar\Omega$ that
has to be assessed~\cite{Konig:2014hma,Furnstahl:2014hca}.
The underlying many-body problem is translationally invariant and thus
preferably expressed in relative coordinates.
For few-body systems like \isotope[4]{He} it is possible to proceed this way,
which includes an exact evaluation of the four-fermion antisymmetrizer. For
systems with more than four nucleons, it is however computationally more
efficient to antisymmetrize in single-particle
coordinates~\cite{Navratil:1999pw}.
To leverage a comparison between the EC emulator and exact solutions we truncate
the HO basis expansion at $\Nmax=16$ for a frequency $\hbar\Omega=36~\MeV$,
which typically gives sub-percent accuracy for the ground-state energy and
radius of \isotope[4]{He}.
With this choice the HO basis consists of 2775 antisymmetric and translationally
invariant four-body states.

The nuclear potential that we employ is additive in the $d=16$ LECs, \ie, we can
express the Hamiltonian as $H(\vec{c}) = H_0 + \sum_{i=1}^{d} c_i H_i$, where
$H_0$ includes the kinetic energy.
Any Hamiltonian with more than one interaction parameter can be written in this
form, where each $c_i$ in general may depend nonlinearly on other parameters.
Furthermore, each term $H_i$ for $i=1, \dots ,16$ can be projected onto the EC
subspace once and then used for an arbitrary number of emulations.
Each of these corresponds to a straightforward solution of the $\NEC \times
\NEC$-dimensional generalized eigenvalue problem.
Unless \NEC is very small, it can in practice happen quite easily that the
EC subspace contains vectors which are almost linearly dependent, leading to a
nearly singular norm matrix.
This problem can be avoided by running an orthogonalization on the EC vectors
that stabilizes the subsequent numerical steps and reveals the effective
dimension of the EC subspace.
Since this step leads to a unit norm matrix, it also reduces the per-sample
evaluation cost at the price of additional preprocessing effort (see
Appendix~\ref{sec:appendix_cost}).

\begin{figure}[t!]
\centering
\includegraphics[width=0.99\columnwidth]{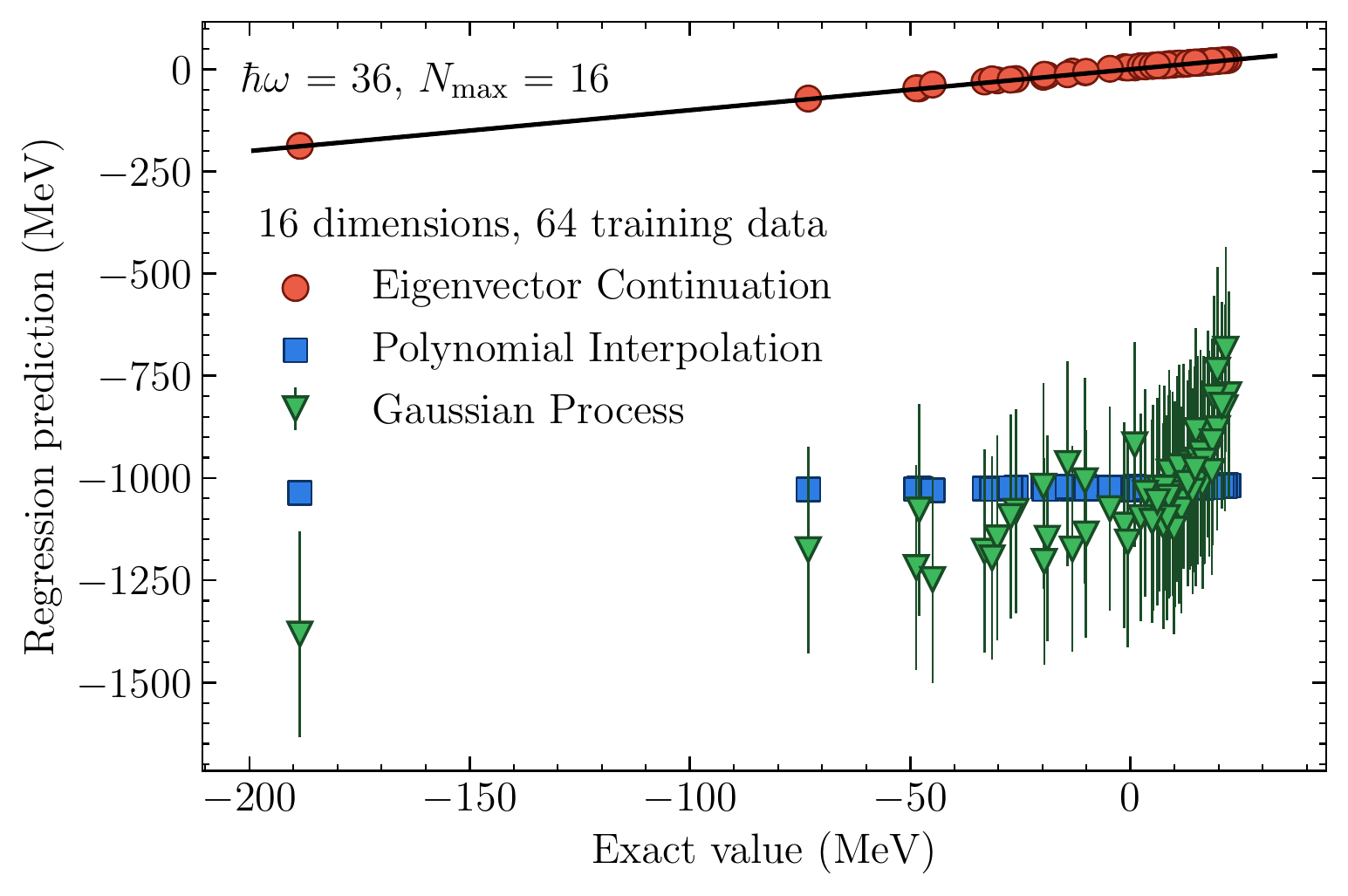}
\caption{%
Comparison of different emulators for the \isotope[4]{He} ground-state energy
using 64 training data points to explore a space where all 16 LECs are varied.}
\label{fig:Xval_4He_E_Nmax16_hw36-pge-d16-n64-int}
\end{figure}

\paragraph*{Results}

To systematically investigate the quality of the EC emulator, we consider
several different cases for the number of LECs that we vary simultaneously,
amounting to sampling Hamiltonians in a $d$-dimensional parameter space, where
$d = 1, \dots , 16$.
We select the set of training points $T = \{\vec{c}^{[i]}\}_{i=1}^{\NEC}$ using
a space-filling Latin Hypercube design~\cite{McKay:1979xx}.
For simplicity we define a parameter domain for each LEC between ${-}2$ and $2$
in appropriate units of inverse energy, see, \eg, Ref.~\cite{Ekstrom:2015rta}.
Validation data is drawn randomly from a uniform distribution
$\mathcal{U}(-2,+2)$.
Each validation point $\vec{c}$ corresponds to either interpolation or
extrapolation from the set of training points, with the former being defined as
the case where $\vec{c}$ lies within the convex hull of $T$.
By randomly generating a coefficient vector $\vec{\alpha}$ with $\alpha_k\geq0$
for $k=1, \dots ,d$ and $\sum_k \alpha_k=1$ it is possible to alternatively
sample only points $\sum_k \alpha_k \vec{c}^{[k]}$ corresponding to
interpolation.
We present results as a cross-validation plots where we consider emulated values
as a function of the exact ones.
In these plots we include results for polynomial interpolation and a Gaussian
process for comparison.
The Gaussian process is constructed using a standard squared exponential kernel
with hyperparameters estimated from the maximum of the marginal log-likelihood
of the calibration data.
A Python script able to run calculations of this type is provided as
Supplemental Material along with this Letter, with a brief description given
in the appendix.

\begin{figure}[t!]
\centering
\includegraphics[width=0.99\columnwidth]{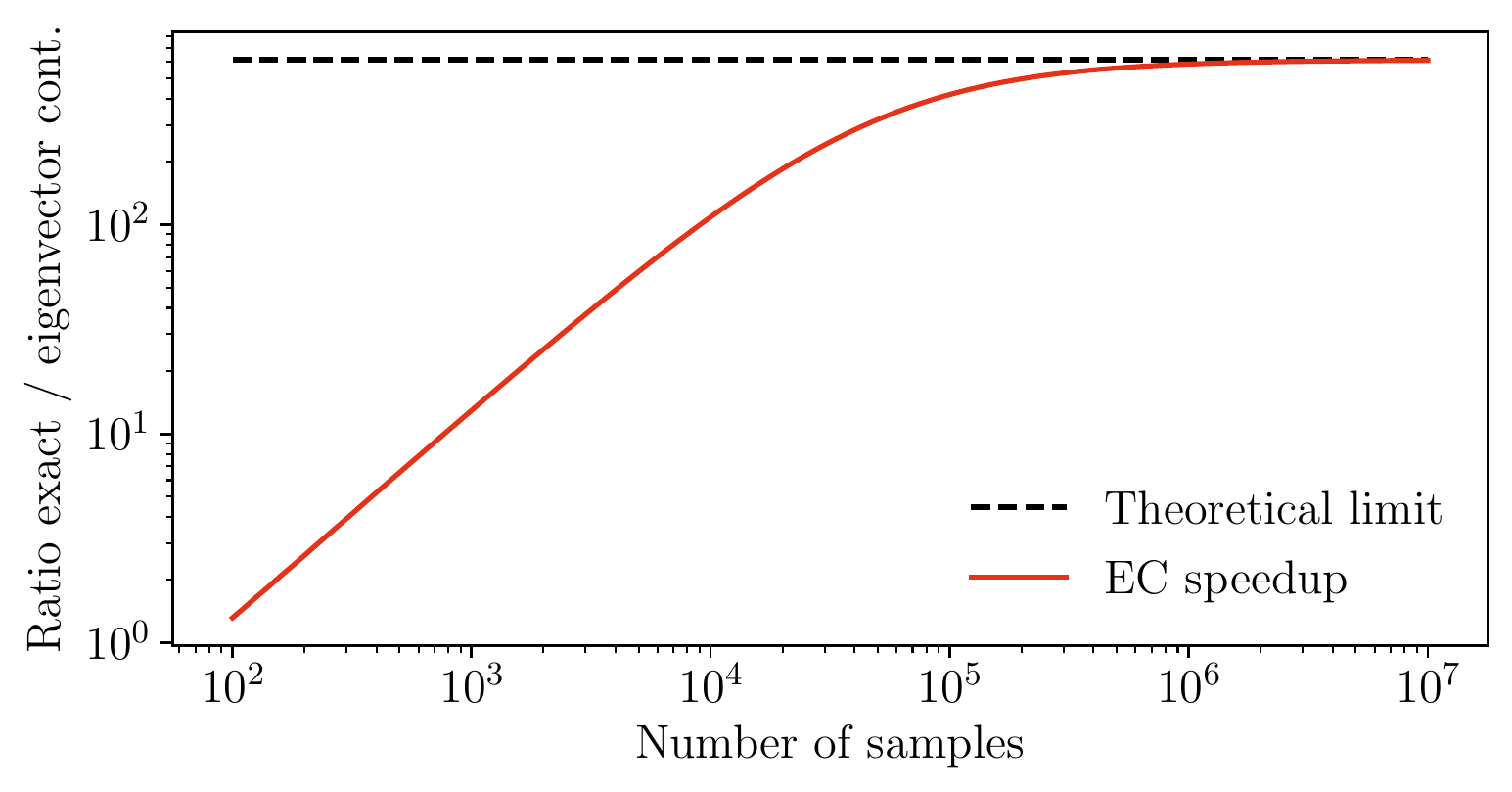}
\caption{Speedup factor (ratio of estimated required floating-point
operations) of EC emulation compared to direct calculation as
function of the number of samples, \ie, number of calls to the
emulator.
The curve shows the result corresponding to the setup as
in Fig.~\ref{fig:Xval_4He_E_Nmax16_hw36-pge-d16-n64-int}, \ie,
varying 16 LECs and using an EC subspace constructed from 64
training data points.
The assumed number of matrix-vector products required for a Lanczos
diagonalization in the full $\Nmax=16$ space is $\NMV=80$ for this case (see
appendix and main text for details).
The theoretical limit indicates the max
speedup reached asymptotically in the number of samples, which is
614 in the present case.}
\label{fig:Speedup_4He_E_Nmax16-d16-n64}
\end{figure}

A representative example is shown in
Fig.~\ref{fig:Xval_4He_E_Nmax4_hw28-pge-d3-n12}.
In this case, calculations for the \isotope[4]{He} ground-state energy are
emulated as a function of three LECs using 12 training data points obtained in
an $\Nmax=16$, $\hw=36~\MeV$ NSCM model space.
Eigenvector continuation is seen to work exceptionally well (the difference to
exact calculations for each point is negligibly small and cannot be resolved in
the plot), whereas polynomial interpolation and the Gaussian process struggle to
provide accurate results even when we consider only validation points
corresponding to interpolation within the convex hull of the set of training
points (right panel in Fig.~\ref{fig:Xval_4He_E_Nmax4_hw28-pge-d3-n12}).

In fact, EC can achieve excellent results even with fewer than 12 training data
points in this particular case.
Furthermore, EC requires only a moderate increase in the number of training data
as the dimension of the parameter space is increased.
In Fig.~\ref{fig:Xval_4He_E_Nmax16_hw36-pge-d16-n64-int} we show results for the
\isotope[4]{He} energy with all 16 LECs varied, using the same $\Nmax=16$,
$\hw=36~\MeV$ NSCM model space as before.
It is evident how EC can still provide accurate results while polynomial
interpolation and the Gaussian process fail completely to emulate the data, even
though only interpolation is considered in
Fig.~\ref{fig:Xval_4He_E_Nmax16_hw36-pge-d16-n64-int}.

To fully appreciate the efficiency gain provided by the EC method, it is
important to compare the overall computational cost of the different methods
considered above.
The cost of emulating with EC is not severe because all relevant matrix
operations, \ie, setting up the target Hamiltonian and solving a generalized
eigenvalue problem, need only be performed in the small EC subspace.
Besides the requirement of carrying out $\NEC$ exact calculations there is a
one-time cost of matrix-matrix-matrix multiplications coming from projecting the
Hamiltonian to the EC subspace.
Thus, the benefit of emulating with EC will improve with the number of calls to
the emulator.
Asymptotically in the number of emulator calls, the speedup of using EC is
proportional to $(M/\NEC)^2$, where $M$ is the dimensionality of the full-space
problem.
Typically, we find $\NEC \approx 10-100$ for problems with $M\approx 10000$,
thus easily yielding a speedup factor $\sim 10^4$ or more.
In Fig.~\ref{fig:Speedup_4He_E_Nmax16-d16-n64} we show the speedup we achieved
for the \isotope[4]{He} problem benchmarked here.
In this particular case the maximum speedup is limited to ``only'' a factor 614,
stemming from the still comparatively small model space that suffices for the
\isotope[4]{He} calculation; it grows rapidly once one considers heavier nuclei.
A detailed analysis of the computational cost is provided in the appendix.

\begin{figure}[t!]
\centering
\includegraphics[width=0.95\columnwidth]{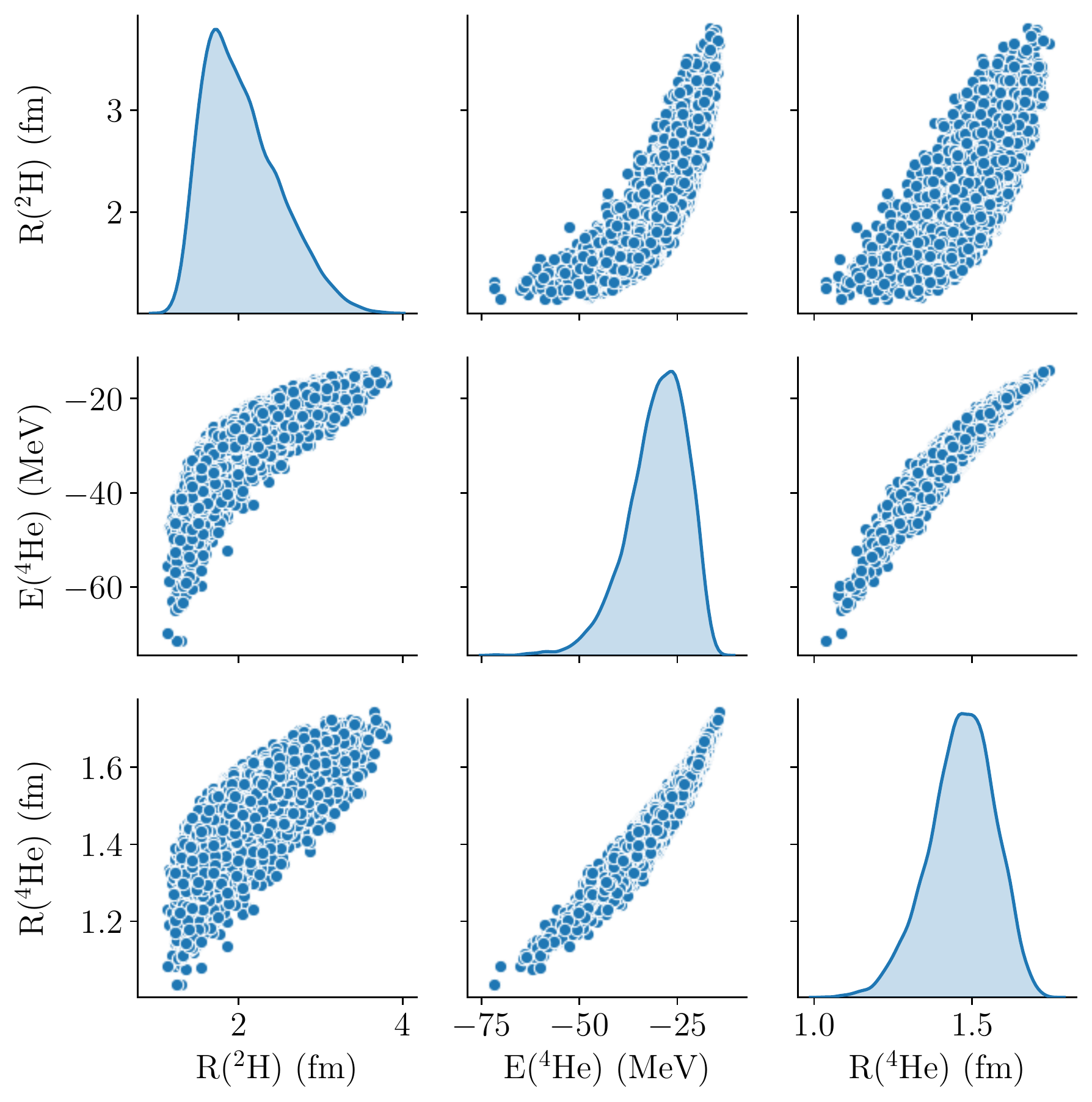}
\caption{%
  Energy-radius correlation between \isotope[2]{H} and \isotope[4]{He}
  based on $10^4$ different values of the 16 LECs that govern
  the NN+3N interaction at NNLO.
  One-dimensional distributions of the emulated values for each observable are
  shown on the diagonal.
  The LECs were varied within a 10\% of the nominal
  NNLO$_{\text{sat}}$~\cite{Ekstrom:2015rta} values.
  The set of panels in the figure is symmetric with respect to the diagonal.
  Remarkably, the entire set of EC evaluations takes less than one minute on a
  standard laptop, a 100-fold speedup compared to exact calculations.
 }
\label{fig:few_body_grid_subset}
\end{figure}

With EC emulation we can efficiently sample all nuclear observables accessible
by, e.g., the NCSM method across a relevant domain of LEC values with
unprecedented efficiency.
In Fig.~\ref{fig:few_body_grid_subset} we present a proof-of-principle
application by correlating selected observables in \isotope[2]{H} and
\isotope[4]{He} across $10^4$ LEC samples at
NNLO~\cite{Carlsson:2015vda,Ekstrom:2015rta}.
Without EC emulation this would be an expensive analysis due to the
large number of three- and four-body calculations.
The significance of EC emulation increases dramatically for heavier nuclei and
enables important but otherwise prohibitively expensive
studies~\cite{Ekstrom:2019lss}.
The known energy-radius correlation in \isotope[4]{He} stands out.
The results also indicate that the radius of \isotope[2]{H} only sets a lower
bound on energy and radius of the \isotope[4]{He}.
This type of study shows the complementary information content in
different observables.
Additional observables, including \isotope[3]{H}, as well as further details
about this analysis are provided in the appendix.

\begin{figure}[t!]
\centering
\includegraphics[width=0.99\columnwidth]{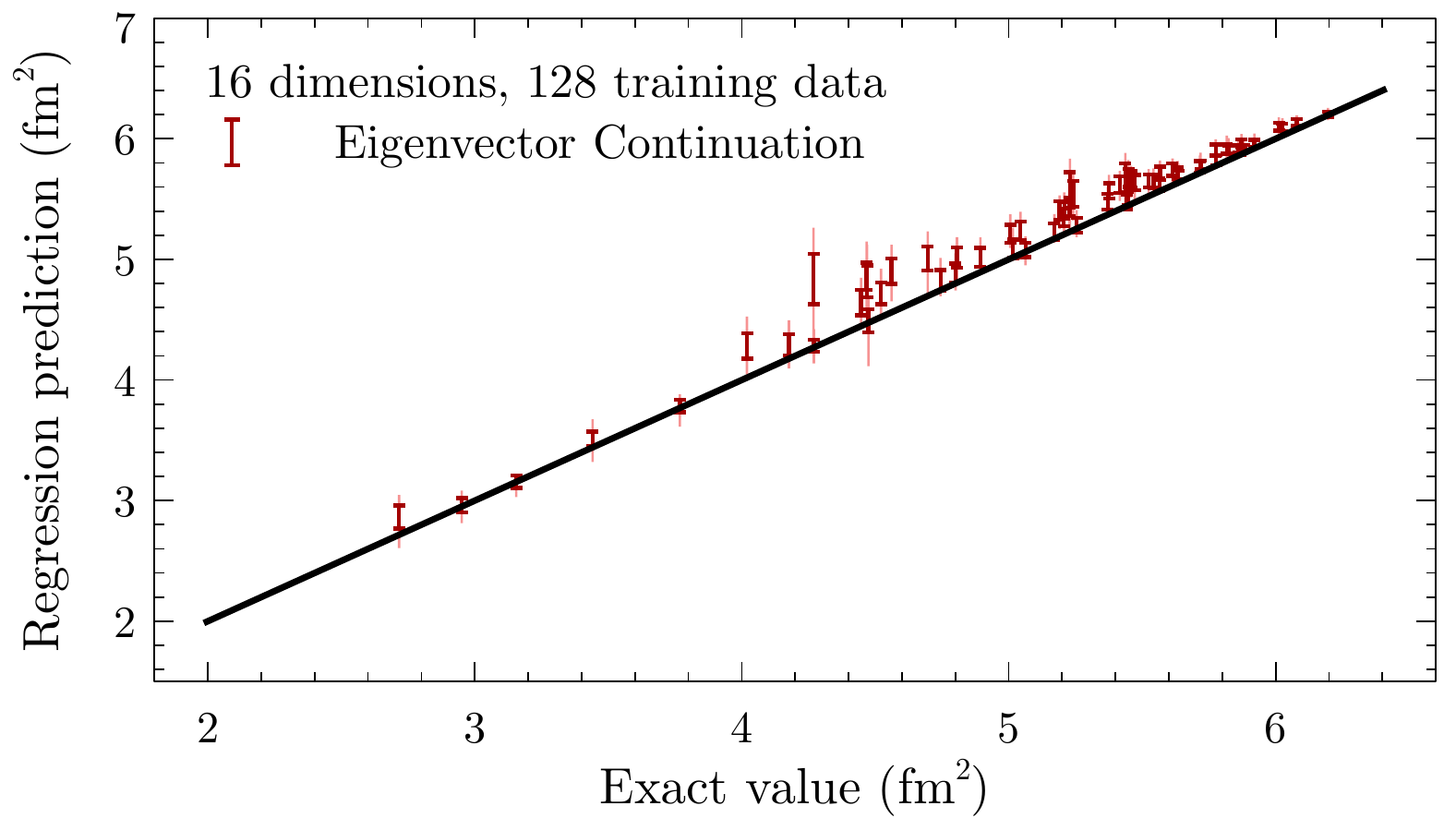}
\caption{%
Cross validation for the \isotope[4]{He} ground-state radius squared using 128
training data points to explore a space where all 16 LECs are varied.
The thicker uncertainty bars indicate 68.2\% intervals obtained by considering
distributions obtained from 32 additional training data sets in addition to the
original sample, while the faint thinner ones indicate the full range of results
for each point.
}
\label{fig:Xval_4He_r2_Nmax16_hw36-e-d16-bf}
\end{figure}

Although GP interpolation cannot deliver the observed
sub-percent accuracy of EC emulation, the GP method provides an uncertainty
estimate of the output value.
For EC, a first detailed analysis of its rate of convergence as the number of
training points is increased has been performed in Ref.~\cite{Sarkar:2020mad}.
However, even without a fully developed theory for EC uncertainties, we can make
the following remarks:
First, EC is a variational method.
This can be seen directly by noting that it is based on constructing a subspace:
considering the original Hamiltonian in diagonalized form, it is clear that
removing any of the basis vectors can only increase the lowest eigenvalue of the
remaining operator.
Therefore, EC-emulated (energy) eigenvalues will always be larger than or equal
to the true result, \ie, resulting in one-sided error bars.
Note, however, that this argument does not apply to other operators evaluated in
the EC subspace.
Second, the fact that EC provides remarkably accurate results with only a small
amount of training data, as well as the benefit that it can reliably both
interpolate and extrapolate, can be exploited in order to estimate the
uncertainty based on removing different points from the training set,
giving a range of values for each emulation target point.
For many applications, this may be an efficient strategy to assess how converged
the EC-emulated values are.
For a more thorough analysis one can use various training sets of the same size
and analyze the distribution of results.  We have used this strategy to obtain
the results for the \isotope[4]{He} ground-state radius squared shown in
Fig.~\ref{fig:Xval_4He_r2_Nmax16_hw36-e-d16-bf}.
Specifically, the figure shows uncertainty bars obtained by considering 32
additional training data sets of 128 points each.
The thicker uncertainty bars correspond to 68.2\% intervals obtained from the
distribution of these results, while the faint thinner ones indicate the full
range of results for each point.
The results indicate that the size of the resulting uncertainty bars correlates
well with the degree of deviation from the exact results and hence serve as a
possible reasonable estimate for the uncertainties.

\paragraph*{Conclusion and outlook}

We have demonstrated how EC can be used to construct an efficient and accurate
emulator of eigenvalue problems with continuous and high-dimensional parametric
dependencies.
Moreover, for systems with a matrix representation that linearly depends on a
set of parameters, the EC method enables a substantial computational speedup
while maintaining high-accuracy outputs compared to exact solutions of the
original problem.
This is achieved by constructing a tailored low-dimensional subspace spanned by
exact eigenvectors for a set of ``training'' points in the parameter space.
We constructed an efficient and accurate emulator of the quantum-mechanical
solution of the \isotope[4]{He} nucleus, considering its ground-state and
squared radius as concrete observables.
The computational speedup offered by the EC emulator is essential for sampling
high-dimensional regions in the parameter domain of any model with the purpose
of, \eg, optimization and uncertainty quantification, where the required large
number of exact calculations would be prohibitively expensive.
For nuclear physics, the EC method can be a key ingredient to facilitate
large-scale Markov-Chain Monte Carlo evaluations of relevant Bayesian posteriors
of the parameters in EFTs or models of the nuclear forces.
Applications to this and related studies are already under way.

\begin{acknowledgments}
This work was supported by the Deutsche Forschungsgemeinschaft (DFG, German
Research Foundation) -- Projektnummer 279384907 -- SFB 1245, the U.S.\
Department of Energy (DE-SC0018638 and DE-AC52-06NA25396), and the European
Research Council (ERC) under the European Union’s Horizon 2020 research and
innovation programme (Grant agreement No. 758027).
This material is based upon work supported by the U.S.\ Department of Energy,
Office of Science, Office of Nuclear Physics, under the FRIB Theory Alliance
award DE-SC0013617.
We thank the Institute for Nuclear Theory at the University of
Washington for hospitality during program INT 19-2a \emph{Nuclear Structure
at the Crossroads}.
\end{acknowledgments}

\appendix

\section{Cost comparison}
\label{sec:appendix_cost}

For the following analysis, we let $M = M(\Nmax)$ denote the actual dimension
of the model space considered in a given calculation (suppressing the
dependence on the number of nucleons). Furthermore, $\NEC$ is the number of
training data points, \ie, the number of states spanning the EC
subspace, while $N$ denotes the number of requested samples. We assume that
sufficient memory is available to store intermediate results as necessary and
we limit the analysis to basic estimates for the required operations, not
taking into account specific optimizations that may be used in practice.
\begin{itemize}
\item We first consider the cost of performing a single calculation in the full
$M$-dimensional space.
Setting up the Hamiltonian, given by a part independent of LEcs
plus a linear combination of terms for each individual LEC, $H = H_0 +
\sum_{\alpha=1}^{\NLEC} c_\alpha H_\alpha$, costs a total of $2\NLEC M^2$
floating-point operations.
Subsequently calculating the ground-state energy
with a Lanczos-like algorithm has a complexity that is dominated by performing
$\NMV$ $M$-dimensional matrix-vector multiplications, each of which costs
$M^2$ operations.
Note that the specific value of $\NMV$ depends on the
desired accuracy of the calculation as well as on the properties of the
Hamiltonian.
In particular, $\NMV$ typically grows with increasing $M$.
Neglecting other aspects of the diagonalization procedure, we arrive at a
total cost of $M^2 \times (2\NLEC + \NMV)$ operations.
\item Multiplying the above by $N$ gives the cost for a direct sampling within
the full space.
\item Setting up an emulator has a base cost of $\NEC \times M^2 \times (2\NLEC
+ \NMV)$.
For polynomial interpolation and Gaussian process emulation we take
this as the total cost and assume the subsequent cost for obtaining samples is
negligible.
\item Setting up sampling based on EC requires some
additional work.
\begin{enumerate}
\item  Given the training set $\{\vec{c}_i\}_{i=1}^{\NEC}$, calculating the norm
matrix involves (neglecting symmetry) $\NEC^2$ $M$-dimensional vector-vector
products, amounting to a cost of $2\NEC^2M$ operations.
\item Similarly, reducing the individual Hamiltonian terms to the training
subspace costs $\NEC$ $M$-dimensional matrix-vector multiplications plus another
$\NEC^2$ vector-vector multiplications, amounting to a total cost of $(\NLEC +
1) \times (2\NEC M^2 + 2\NEC^2M)$.
\end{enumerate}
\item For each point sampled using EC, the Hamiltonian setup then only needs to
be performed in the subspace, amounting to $2\NLEC \NEC^2$ operations per
sample.
Solving the generalized eigenvalue problem costs another $14 \NEC^3$
operations~\cite{Golub:1996}.
\item The sampling cost can be reduced by performing an initial
orthogonalization of the $\{\vec{c}_i\}_{i=1}^{\NEC}$ (which we assume to be
achieved through a singular-value decomposition costing about $6M\NEC^2
+ 11\NEC^3$ operations~\cite{Golub:1996}), leaving only the solution of a
standard symmetric eigenvalue problem and thus a cost of $26\NEC^3/3
+ \OO(\NEC^2)$ operations per sample~\cite{Demmel:1997}.
\end{itemize}

\section{Emulating the relation between selected few-body observables}

\begin{figure*}[t!]
\centering
\includegraphics[width=0.59\textwidth]{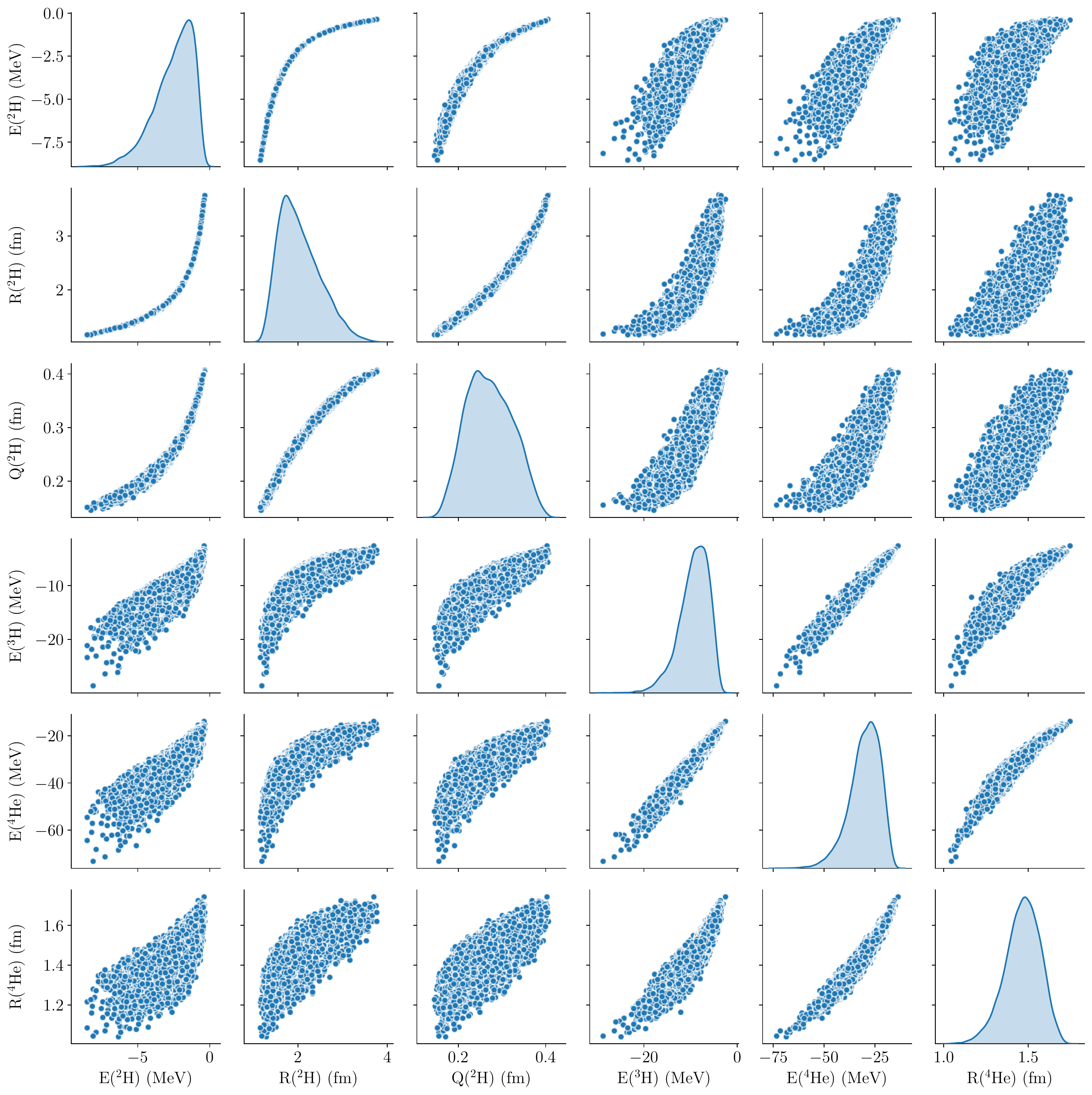}
\caption{%
  Emulated sample of how selected observables in
  \isotope[2]{H}, \isotope[3]{H}, and\isotope[4]{He} co-vary with
  each other at $10^4$ different values of the 16 LECs that govern the
  NN+3N interaction at NNLO.
  The set of panels in the figure is symmetric with respect to the diagonal.
  Remarkably, the entire set of EC evaluations less than two minutes on a
  standard laptop.
 }
\label{fig:few_body_grid}
\end{figure*}

Figure~\ref{fig:few_body_grid} presents the results from sampling the mutual
co-dependence of a larger set of observables in \isotope[2]{H}, \isotope[3]{H},
and \isotope[4]{He}.
The LECs were drawn using latin hypercube sampling within a parameter domain
defined from a 10\% variation of the nominal NNLO$_{\text{sat}}$ values.
This represents a typical size of a relevant parameter domain of a chiral
interaction.
Given the small numerical value of the LEC $c_E \approx -0.04$ in
NNLO$_{\text{sat}}$, the limit of this particular LEC is multiplied with a
factor of 20.
This makes the resulting domain limit in this direction comparable with the
other LEC limits.

As expected, the results from the sampling show a strong correlation between
all observables in \isotope[2]{H}.
The same observation can be made between the observables in \isotope[3]{H} and
\isotope[4]{He}.
The lack of correlation between $A=2$ and $A=3,4$ observables clearly indicates
the effect of the $3N$ interaction that is not present in the $A=2$ system.

\section{Python code}

We provide the Python code \texttt{ec\_xval.py} as Supplemental Material.
This program is a simplified version of the script that was used to generate
the cross-validation plots shown in the main text.
Matrices required as input data (NCSM Hamiltonian along with corresponding
representation of radius squared operator) are provided as well.
Due to storage limitations, these matrices are restricted to rather small
NCSM model spaces, but they nevertheless provide representative examples.
It is our hope that this code will facilitate applications of eigenvector
continuation to a variety of cases where efficient and accurate emulators
are required.
Making use of freely available Python packages, the code generates
cross validation plots that compare EC to both a Gaussian process and simple
polynomial interpolation.

Running a cross validation is as simple as typing
\begin{verbatim}
$ python3 ex_xval.py -d 3 -n 6
\end{verbatim}
in the terminal.
This will generate a cross-validation plot for a three-dimensional parameter
space, using 6 EC basis vectors.
By default, the cross validation is run using only eigenvector continuation.
In order to compare at the same time to polynomial interpolation and a Gaussian
process, as shown in the main text, \texttt{-pge} can be given as an option to
enable all emulators.
A number of further aspects can be controlled by passing command-line options,
a full list of which, along with explanations, is printed to the terminal by
running:
\begin{verbatim}
$ python3 ex_xval.py --help
\end{verbatim}

\end{document}